\documentstyle[12pt]{article}

\newcommand{\dd}{\partial}
\newcommand{\De}{\Delta}
\newcommand{\f}{\varphi}
\newcommand{\ls}{\left(}
\newcommand{\rs}{\right)}
\newcommand{\g}{\gamma}
\newcommand{\m}{\mu}
\newcommand{\n}{\nu}
\newcommand{\ra}{\rangle}
\newcommand{\te}{\theta}

\newcommand{\disn}[2]{$$\displaylines{\refstepcounter{equation}%
            \label{#1}\hskip 1em minus 1em #2\hfilneg}$$}
\newcommand{\nom}{\hfil\hskip 1em minus 1em (\theequation)}

\makeatletter

\newcount\dobav
\newcount\otlog
\newcount\vrem
\def\@citex[#1]#2{\if@filesw\immediate\write\@auxout{\string\citation{#2}}\fi
  \let\@citea\@empty
  \dobav=-1
  \otlog=-1
  \@cite{\@for\@citeb:=#2\do
    {\def\@tempa##1##2\@nil{\edef\@citeb{\if##1\space##2\else##1##2\fi}}%
     \expandafter\@tempa\@citeb\@nil
     \@ifundefined{b@\@citeb}{\@warning%
       {Citation `\@citeb' on page \thepage \space undefined}%
       \vrem=-1}{\vrem=\csname b@\@citeb\endcsname}
\advance\vrem by -1 \ifnum \vrem=\dobav
 \otlog=\vrem
 \advance\otlog by 1
\else
 \ifnum \vrem=\otlog
  \advance\otlog by 1
 \else
  \ifnum \otlog>0
   \advance\dobav by 1
   \ifnum \otlog=\dobav
    \hbox{,\penalty\@m\ \the\otlog}%
   \else
    \hbox{--\the\otlog}%
   \fi
   \otlog=-1
  \fi
  \dobav=\vrem
  \advance\dobav by 1
  \@citea\def\@citea{,\penalty\@m\ }%
  \ifnum \dobav=-1
   {\reset@font\bf ?}%
  \else
   \hbox{\the\dobav}%
  \fi
 \fi
\fi
}%
\ifnum \otlog>0
 \advance\dobav by 1
 \ifnum \otlog=\dobav
  \hbox{,\penalty\@m\ \the\otlog}%
 \else
  \hbox{--\the\otlog}%
 \fi
\fi }{#1}}

\renewcommand{\section}{\@startsection{section}{1}{0pt}%
          {3.5ex plus 1ex minus .2ex}{2.3ex plus .2ex}{\noindent\hfil\bf}}

\makeatother

\newcommand{\st}{\protect\\ $\protect\vphantom{0}$\hfil}

\begin{document}

\title{
Ultraviolet finiteness of Chiral Perturbation Theory\\
for two-dimensional Quantum Electrodynamics\\}
\author{S.~A.~Paston\thanks{E-mail: paston@pobox.spbu.ru},
E.~V.~Prokhvatilov\thanks{E-mail: Evgeni.Prokhvat@pobox.spbu.ru},
V.~A.~Franke\thanks{E-mail: franke@pobox.spbu.ru}\\
St.-Petersburg State University, Russia}
\date{\vskip 15mm}

\maketitle

\begin{abstract}
We consider the perturbation theory in the fermion mass (chiral
perturbation theory) for the two-dimensional quantum
electrodynamics. With this aim, we rewrite the theory in the
equivalent bosonic form in which the interaction is exponential
and the fermion mass becomes the coupling
constant. We reformulate the bosonic perturbation theory in the
superpropagator language and analyze its ultraviolet behavior. We
show that the boson Green's functions without vacuum loops remain
finite in all orders of the perturbation theory in the fermion
mass.
\end{abstract}

\newpage

\section{Introduction}
Because of the difficulties that arise in analyzing gauge
theories, it is interesting to investigate simple models that
admit a nonperturbative description and, in particular, to
investigate infinite perturbation theories (PT) in all orders.
One such model is the two-dimensional quantum electrodynamics
(2D-QED) with a nonzero fermion mass. This model, also called the
massive Schwinger model, is described by the Lagrangian density
 \disn{1}{
 L=-\frac{1}{4}F_{\m\n}F^{\m\n}+\bar\Psi(i\g^{\m}D_\m-M)\Psi,
 \nom}
where $\m,\n=0,1$,\ \ $F_{\m\n}=\dd_{\m}A_{\n}-\dd_{\n}A_{\m}$, \
$D_{\m}=\dd_{\m}-ie_{\rm el}A_{\m}$,\ \  $A_{\m}$
is the vector potential of the electromagnetic field,
\ \
$\Psi = {\psi_- \choose \psi_+}$,\ \ \  $\bar\Psi =\Psi^+\g^0$
is the field of the fermion with the mass
$M$,\ \  $e_{\rm el}$ is the analogue of the electron charge, and the matrices
$\g^{\m}$ can be taken in the form
 \disn{1.1}{
\g^0=\ls
\begin{array}{cc}
0 & -i\\
i & 0
\end{array}
\rs, \quad \g^1=\ls
\begin{array}{cc}
0 & i\\
i & 0
\end{array}
\rs.
\nom}
It is known that this model is exactly solvable \cite{tochn} in the case
of a vanishing fermion mass $M$. In the case of $M\ne  0$, we can
therefore consider the so-called chiral PT, i.e., the PT in the
fermion mass $M$ \cite{adam}. This is interesting because such a PT
corresponds to an approach that is nonperturbative from the
viewpoint of the usual PT. Indeed, the actual expansion is in
the dimensionless ratio $M/e_{\rm el}$ (the coupling constant $e_{\rm el}$
has the
dimension of mass for a two-dimensional theory). We can therefore
consider the chiral PT as an expansion in the parameter $1/e_{\rm el}$,
which is valid in the domain of large $e_{\rm el}$.

In gauge theories with space-time dimensionality greater than two, the
vanishing fermion mass does not suffice for making the model
exactly solvable; therefore, the PT in the parameter $M$ cannot be
constructed. One can nevertheless try to consider the expansion in
the parameter $1/e_{\rm el}$ in such theories, passing to the "dual"
form of the model. We can study the features of such expansions using
the chiral PT for 2D-QED, which provides a simple example of it.

One more domain of application of the chiral PT for 2D-QED is
constructing the light-front (LF) Hamiltonian for such a model
and studying the properties of the such approach on
this example. The LF Hamiltonian approach \cite{dir} is completely
nonperturbative (i.e., the corresponding Schrodinger equation can
be solved at arbitrary values of the coupling constant), while
the description of the physical vacuum state becomes trivial in
this approach \cite{cla}. But because of the breaking of Lorentz
invariance and the presence of singularities specific to the
LF coordinates at the zero value of the lightlike momentum
$p_-=(p_0-p_3)/\sqrt{2}$\ \ \cite{nov1,nov2,nov3},
the theory generated by the
LF Hamiltonian can be nonequivalent to the initial
Lorentz-covariant theory \cite{bur,naus}. In this case, restoring the
equivalence requires adding coun-terterms to the LF Hamiltonian.
The form of these counterterms can be found either by using the
method of approximate limiting transition to the LF \cite{naus,pred1,pred2}
or
by comparing the contributions in all orders of the PT in the
coupling constant for the theory generated by the LF Hamiltonian
and for the Lorentz-covariant theory \cite{bil,tmf97,tmf99}. But the latter
method cannot be applied to 2D-QED, because the PT in the
coupling constant $e_{\rm el}$ fails in this model due to infrared
divergences. Nevertheless, it is possible to construct the LF
Hamiltonian for 2D-QED using the chiral PT \cite{shw1,shw2,tmf02}.

The chiral PT for 2D-QED can be conveniently constructed by
passing to the equivalent formulation of the theory in the
bosonic variables using the bosonization procedure \cite{colm,prok,naus}.
The corresponding bosonic Lagrangian is
 \disn{2}{
L=\frac{1}{8\pi}\ls\dd_\m\f\dd^\m\f-m^2\f^2\rs+
\frac{\g}{2}e^{i\te}:e^{i\f}:+\frac{\g}{2}e^{-i\te}:e^{-i\f}:,\quad
 \g=\frac{Mme^C}{2\pi},\quad
 m=\frac{e_{\rm el}}{\sqrt\pi},
 \nom}
where $C=0.577216$ is the Euler constant, the parameter $\te$
is responsible for the choice of the instantonic $\te$- vacuum in the 2D-QED
\cite{colm,prok,adam} and the normal-ordering symbol means that the PT in
$\g$ does not contain diagrams with loops containing only one
vertex (this corresponds to the standard normal-ordering symbol
in the Hamiltonian).

This Lagrangian describes the theory of a massive scalar field
with an exponential interaction. The fermion mass $M$ then plays
the role of the coupling constant, and the chiral PT for 2D-QED
therefore coincides with the standard PT for bosonic theory (\ref{2}).
The propagator is simple in this PT (it differs from the standard
scalar field propagator only by a numerical factor), but
Lagrangian (\ref{2}) contains a nonpolynomial interaction. In each
order of the PT in $M$, we therefore have an infinite number of
diagrams. Each separate diagram converges, but it turns out that
particular infinite sums of diagrams may diverge in a given order
of the PT in $M$, and this divergence manifests the ultraviolet
(UV) nature \cite{adam,shw1}. The question of the presence of UV
divergences in the PT in $M$ for bosonic Green's functions is
therefore nontrivial, although two-dimensional models are usually
free of UV divergences. We note that this problem was not
been completely solved \cite{adam,adamch}.

In this paper, we prove the absence of UV divergences in all
orders of the PT in $\g$ (and hence in $M$) for the Green's functions
without vacuum loops in bosonic theory (\ref{2}). We understand the
absence of UV divergences to mean the finiteness of the result in
the limit of removed UV regularization (we need an intermediate
UV regularization because of the absence of absolute convergence
of the sum of diagrams in a given order of the PT expansion in
$\g$). The result is proved to be finite because all divergences
cancel when all the contributions of the given order in $\g$ are
summed in any given Green's function.

\section{Absence of surface UV divergences in Green's functions\st
of orders higher than the second in {\protect\large $\g$}}

We consider the structure of diagrams of the Feynman PT in $\g$ for
Lagrangian (\ref{2}). There are two types
of vertices with $j$ external lines ($j=0,1,2,\dots$)
generated by two interaction terms in (\ref{2}). The factors
 \disn{3}{
 i^{j+1}\,\frac{\g}{2}e^{i\theta} \qquad {\rm and}\qquad
 i^{-j+1}\,\frac{\g}{2}e^{-i\theta}
 \nom}
correspond to these vertices respectively for the first and second
types of interaction. Vertices with $j=0$, i.e., without lines,
must be considered subdiagrams of a nonconnected diagram. It is
convenient to relate the part $i^{\pm j}$ of the vertex factors
to the lines
that are external with respect to a vertex (we set $\pm i$ to each
line); the vertex factors then become
 \disn{3.2}{
 i\,\frac{\g}{2}e^{i\theta} \qquad {\rm and}\qquad
 i\,\frac{\g}{2}e^{-i\theta}.
 \nom}
The propagator $\Delta(x) = \langle 0|T\f(x)\,\f(0)|0\ra$,
where the field $\f(x)$
corresponds to the free theory described by the Lagrangian
(\ref{2}) with $\g=0$, is
 \disn{4}{
 \De(x)=\int d^2k\;e^{ikx}\De(k),\qquad
 \De(k)=\frac{i}{\pi}\frac{1}{(k^2-m^2+i0)},
 \nom}
where $d^2k=dk_0dk_1$, $kx=k_0x^0+k_1x^1$.

Because the theory with Lagrangian (\ref{2}) contains exponential
interaction terms, we can reformulate the PT in $\g$ in the
superpropagator language, i.e., in terms of sums of contributions
of the same order in $\g$ corresponding to all possible variants of
joining a pair of vertices by different numbers of propagators
\cite{adam,shw2}. For a pair of vertices of different types, the
superpropagator is
 \disn{5}{
 \sum_{m=0}^{\infty}\frac{1}{m!}\De(x)^m=e^{\De(x)},
 \nom}
while for a pair of vertices of the same type, it is
 \disn{6}{
 \sum_{m=0}^{\infty}\frac{1}{m!}(-1)^m\De(x)^m=e^{-\De(x)}.
 \nom}
In formulas (\ref{5}) and (\ref{6}), the factor $1/m!$ is the symmetry
coefficient, which corresponds to $m$ parallel lines. The parts of
vertex factors (\ref{3}) joined to the lines (see the reasoning
after that formula) are taken into account. In such an approach,
the sum of all standard (including nonconnected) diagrams with
the given number of vertices and the given way of attaching the
external line couplings is described by a single diagram in which
each pair of vertices is joined by the corresponding
superpropagator (the connectedness is always understood as the
standard connectedness). For the external lines, we set the usual
propagators (\ref{4}) with the additional factor $\pm i$, which is a part of
the vertex factor related to a line.

We now perform the Wick rotation, thus passing to the Euclidean
space, and introduce an intermediate UV regularization for
propagator (\ref{4}) (e.g., by the higher-derivative method).

We consider the (infinite) sum $S'_n$ of all the diagrams (including
nonconnected) of the order $n$ in the PT in $\g$ for which the
number of external lines coupled to each vertex is fixed and
the type of each vertex is also fixed. In terms of
superpropagators, this sum is a single diagram with each pair of
vertices joined by the corresponding superpropagator (\ref{5}) or (\ref{6}).

We let $S_n$ denote the sum of all connected diagrams
contained in $S'_n$. It can be written in the form
 \disn{6.1}{
 S_n=S'_n-S''_n,
 \nom}
where $S''_n$ is the sum of the corresponding nonconnected diagrams.
The quantity $S''_n$ can be represented as a sum, each term of which
is the sum of all nonconnected diagrams with the fixed partition
of the initial diagram into connected parts. Each such term is
the product of several quantities $S_{\tilde n}$ with $\tilde n<n$.
We can now
perform expansion (\ref{6.1}) for each $S_{\tilde n}$ and repeat this reasoning until
all the obtained quantities of type $S_{\tilde n}$ coincide with $S_1$.
Taking
into account that $S_1=S'_1$ (because there are no nonconnected
diagrams in the first order in PT), we conclude that the quantity
$S_n$ can be represented as a finite sum of products of the
quantities $S'_j$ with $j\le n$:
 \disn{6.2}{
S_n=\sum\prod_kS'_{j_k},
\nom}
 \disn{6.3}{
\sum_kj_k=n.
\nom}

We now investigate the surface UV divergence of the quantity $S_n$
(in the limit of removing the intermediate regularization). The
surface UV divergence is understood to be the degree of
divergence determined by the UV divergence index of a diagram
without taking possible subdiagram divergences into account. When
evaluating this index, we assume that all internal momenta of the
diagram tend to infinity. In the coordinate space, this
corresponds to the case where the coordinates of all the vertices
tend to each other. If we then obtain a pole of order $r_n$ under
the integral sign, then the condition for the absence of the
surface divergence for the quantity $S_n$ is
 \disn{11}{
 2(n-1)-r_n>0,
 \nom}
where we take $n-1$ two-dimensional volume elements into account
and leave one volume element aside because of the translational
invariance.

To estimate the pole order $r_n$, we expand the quantity $S_n$ using
formula (\ref{6.2}), assuming this formula to be applied to expressions
under the integral sign, while we take all integration signs
outside the summation sign. We now find the order $r'_j$ of the pole
that appears when all the coordinates of all the vertices in the
quantity $S'_j$ tend to each other. It follows easily from the
form (\ref{4}) of the propagator that
 \disn{7}{
\De(x)\sim \log\frac{1}{x^2},\qquad {\rm at} \quad x\to 0.
\nom}
Therefore each superpropagator joining vertices of different types
produces a second-order pole when the coordinates of
these two vertices tend to each other (see (\ref{5})),
 \disn{8}{
e^{\De(x)}\sim\frac{1}{x^2},\qquad {\rm at} \quad x\to 0,
\nom}
while each superpropagator joining vertices of the same type
produces a second-order zero (see (\ref{6})),
 \disn{9}{
e^{-\De(x)}\sim x^2,\qquad {\rm at} \quad x\to 0.
\nom}
We now represent the integrand in the quantity $S'_j$ in terms of the
superpropagators, as described after formula (\ref{6}). Because we join
each pair of vertices by a superpropagator of one or the other
type, we can use asymptotic expressions (\ref{8}) and (\ref{9})
to find easily
the total asymptotic expression in the case where all the
coordinates of all the vertices tend to each other, i.e., to
find $r'_j$.

Let $S'_j$ contain $l_1$ vertices of the first type and $l_2$ vertices of
the second type, $l_1+l_2=j$. Then the respective numbers of
propagators connecting vertices of different and the same types
are
 \disn{9.1}{
l_1l_2 \qquad {\rm and} \qquad
\frac{l_1(l_1-1)}{2}+\frac{l_2(l_2-1)}{2}.
\nom}
The overall order of the pole is therefore
 \disn{10}{
r'_j=2\ls l_1l_2-\frac{l_1(l_1-1)}{2}-\frac{l_2(l_2-1)}{2}\rs=
j-(l_1-l_2)^2\le j.
\nom}
Formula (\ref{6.2}) implies that
 \disn{10.1}{
r_n=\max\sum_k r'_{j_k},
\nom}
where the maximum is taken over all summands (\ref{6.2}). Hence, using
estimate (\ref{10}) and condition (\ref{6.3}), we find that
 \disn{10.2}{
r_n\le \max\sum_k j_k=n.
\nom}
Using this inequality, we can estimate the left-hand side of
condition (\ref{11}) for the absence of the surface divergence for the
quantity $S_n$ and thus obtain
 \disn{10.3}{
 2(n-1)-r_n\ge 2(n-1)-n=n-2,
 \nom}
which means that such a divergence is absent for $n>2$. The UV
divergence is clearly absent for $n=1$ (because of the
normal-ordering symbol in Lagrangian (\ref{2}), which prohibits loops
containing only one vertex), and the only case when the surface
divergence exists in the sum of all connected diagrams
of order $n$ in $\g$ is the case $n=2$.

\section{Analyzing divergences in the second order in {\protect\large $\g$}}
At $n=2$, condition (\ref{11}) is broken only if $r_2\ge 2$. By virtue of
formula (\ref{10.1}), this can happen only if $r'_2\ge 2$ because $r'_1=0$ (no
loops containing only one vertex permitted), and condition (\ref{6.3})
must be satisfied. Formula (\ref{10}) then tells us that this is
possible only if $l_1=l_2$; then $r'_2=2$ and hence $r_2=2$. We
therefore conclude that the surface UV divergence can occur only
for the sum of all connected diagrams of the second order in $\g$
with vertices of different types ($l_1=l_2=1$) and with the fixed
arrangement for joining external lines to the vertices. This
divergence turns out to be logarithmic because the left-hand side
of condition (\ref{11}) then vanishes.

The UV-divergent sum of the diagrams just described contributes
to any Green's function in the second order of the PT in $\g$. We
now investigate the presence of the surface UV divergence in the
case of the sum of all such contributions. For the $N$-point
Green's function ($N>0$), the sum of all contributions of the
second order in $\g$ that correspond to the fixed type of vertices
(in the case of vertices of different types) can be found from
Lagrangian (\ref{2}) by virtue of formulas (\ref{3})-(\ref{6}):
 \disn{16}{
 G_N^{(2)}(y_1,\dots,y_N)=-\ls\frac{\g}{2}\rs^2
 \int d^2x_1d^2x_2\prod_{i=1}^N\ls\sum_{k_i=1,2}i\,(-1)^{k_i+1}
 \De(y_i-x_{k_i})\rs e^{\De(x_1-x_2)}.
 \nom}
where the summation ranges over all possible variants of joining
external lines to various internal vertices. As $(x_1-x_2)\to 0$, the
expression in brackets in (\ref{16}) under the product sign behaves as
 \disn{17}{
\sum_{k_i=1,2}i\,(-1)^{k_i+1}\De(y_i-x_{k_i})=
i\Bigl(\De(y_i-x_{1})-\De(y_i-x_{2})\Bigr)
\sim O(x_1-x_2),
\nom}
i.e., it improves the convergence of the integral at the point $(x_1-x_2)=0$.
Because we have just proved that without taking this
into account, the integration in the vicinity of this point
(i.e., when the coordinates of
the vertices tend to each other) can develop at most a logarithmic UV
singularity, we conclude that with asymptotic behavior
(\ref{17}) now taken into account, the $N$-point Green's function in the
second order of the PT in $\g$ does not contain a surface UV
divergence for $N>0$.

Because, as already mentioned, no UV divergences occur in the
first order of the PT in $\g$, the sums of diagrams of the Green's
functions cannot contain divergent subdiagrams in the second
order of the PT. Therefore, because the surface (i.e., not taking
possible subdiagram divergences into account) divergence is
absent, the $N$-point Green's function is actually UV finite in
the second order of the PT in $\g$ for $N>0$.

\section{The absence of subdiagram divergences}
We now address the question whether the quantities $S_n$ with $n>2$
can contain UV divergences because of the presence of UV
divergent sums of subdiagrams.

At $n=3$, only a sum of subdiagrams of the second order in $\g$ with
vertices of different types can be UV divergent (see the
beginning of Sec.~3). Only connected diagrams enter the quantity
$S_n$, and each subdiagram of the second order therefore contains at
least one line that is external with respect to this subdiagram.
In the total sum of all the diagrams constituting the quantity
, we encounter the summation over all possible ways of joining
this line to the subdiagram vertices, which is completely
analogous to the summation over all possible ways of joining
external lines to a diagram of the second order (see Sec.~3).
Using formulas analogous to (\ref{16}) and (\ref{17}), we can then conclude
that the sum of subdiagrams of the second order in $\g$ is UV
finite. Hence, there is no UV divergence for the quantity $S_n$ in
the case $n=3$.

This, in particular, means that only a sum of subdiagrams of the
second order might result in an UV divergence in the case $n=4$,
and we can repeat the above reasoning for this case as well.
Further, using the induction process, we obtain the result: the
sum of all connected diagrams of order $n>2$ in the PT in $\g$ is
actually (not only by counting UV-divergence index) UV finite.
This result holds for diagrams with any number of external lines
and, in particular, for the vacuum diagrams with $n>2$.

Combining this result with the conclusions in Sec.~3, we obtain
the actual UV finiteness of all connected diagrams of the
$N$-point Green's function in all orders of the PT in $\g$ for $N>0$
(i.e., for all diagrams except vacuum diagrams). Hence, we obtain
the UV finiteness of all the Green's functions without vacuum
loops because the diagrams for such Green's functions are
products of connected nonvacuum diagrams.

\section{Conclusion}
For the PT in $\g$ (and hence in the fermion mass $M$) in bosonic
formulation (\ref{2}) of 2D-QED, we have proved the UV finiteness of
the following quantities:

1.~For terms of order $n>2$ of the PT, the sum of all connected
diagrams with any (in particular, zero) number of external lines
is UV finite; this remains valid if we fix the way the external
lines of a diagram are attached to its vertices.

2.~In the second order of the PT, all nonvacuum Green's
functions, i.e., the sums of all diagrams with a nonzero number
of external lines, become UV finite after summing up all the
contributions corresponding to all possible ways of joining the
external lines of the diagram to its vertices.

In particular, we then conclude that Green's functions without
vacuum loops are UV
finite in all orders of the PT in $\g$.

Only the sum of all vacuum diagrams and the sum of all diagrams
with a nonzero number of external lines but with a fixed way of
joining these lines to the vertices of the diagram are UV
divergent in the second order of the PT in $\g$. This divergence is
logarithmic. The vacuum Green's function
(i.~e. without external lines) is therefore
logarithmically UV divergent, but only in the second order of the
PT in $\g$.

Because only Green's functions without vacuum loops are
important, we have thus proved the UV finiteness of the chiral PT
for 2D-QED.

\vskip 1em
{\bf Acknowledgments.}
The work was supported in part by the Russian Federation Ministry
of Education (Grant No. E00-3.3-316).

\end{document}